# On the concept of mass point in general relativity

A. LOINGER

Dipartimento di Fisica, Università di Milano

Via Celoria, 16 – 20133 Milano, Italy

**Summary.** – *The* correct characterization of the concept of mass point in general relativity is a straightforward consequence of the **original** form of solution given by Schwarzschild to the problem of the Einstein field of a material point.

PACS. 04.20 – General relativity: fundamental problems and general formalism – Solutions to equations.

**1**. – It is commonly believed that the concept of mass point in general relativity is implicitly defined by the Kruskal-Szekeres form of solution to the Schwarzschild problem [1]. I think, on the contrary, that the only appropriate definition of the above concept is given by Schwarzschild's **original** form of solution to the homonymous problem.

Here are my arguments.

**2**. – The static solution of the static problem ("Schwarzschild problem") of the Einstein gravitational field generated by a point mass $M$, at rest, is given by the following expression of the space-time interval:

$$(2.1) \quad ds^2 = \left[1 - \frac{2m}{f(r)}\right] c^2 dt^2 - \left[1 - \frac{2m}{f(r)}\right]^{-1} [df(r)]^2 -$$

$$- f^2(r) [d\theta^2 + \sin^2\theta \, d\varphi^2] \quad ;$$

$$(r > 0; \; 0 \leq \theta \leq \pi; \; 0 \leq \varphi < 2\pi) \quad ,$$

where: $m \equiv GM/c^2$; $G$ is the gravitational constant; $f(r)$ is *any regular* function of $r$ such that $ds^2$ is Minkowskian at the spatial infinite [2].

If we put $f(r) \equiv r$, we obtain the well-known **standard** form of solution, which is due to Hilbert, Droste, and Weyl. It is generally called "Schwarzschild solution", but in reality the **actual** Schwarzschild's form of solution [3] follows from (2.1) by putting:

$$(2.2) \qquad f(r) \equiv \left[ r^3 + (2m)^3 \right]^{1/3} ;$$

Schwarzschild's $ds^2$, which holds for $r > 0$, is **diffeomorphic** to the "exterior" part $r > 2m$ of the standard $ds^2$.

Another interesting form, valid for $r > 0$, can be obtained with the choice

$$(2.3) \qquad f(r) \equiv r + 2m ;$$

the corresponding $ds^2$ is diffeomorphic to Schwarzschild's.

**3**. − The elementary interval of Kruskal-Szekeres [1] is, with slight changes of notations,

$$(3.1) \qquad d\sigma^2 = F^2(-dv^2 + du^2) + r^2 (d\theta^2 + \sin^2\theta \, d\varphi^2) ,$$

where $u$ and $v$ are suitable functions of $r$ and $t$, and

$$(3.2) \qquad u^2 - v^2 = \left(\frac{r}{2m} - 1\right)\exp\left(\frac{r}{2m}\right) ;$$

$$(3.3) \qquad F^2 = (32m^3/r)\exp\left(-\frac{r}{2m}\right) = \text{a trascendental function of } u^2 - v^2.$$

From the physical standpoint, formula (3.1) has several drawbacks, as it was pointed out by many authors. I shall emphasize here the following four defects. First of all, (3.1) yields a *non*-static Einstein field as a solution of a static problem: this is rather baroque. And also baroque is the circumstance that each point of the standard form is represented *twice* by the Kruskal-Szekeres form, while the elliptical reduction would encounter insuperable difficulties. Then, the derivatives $\partial u/\partial r$ and $\partial v/\partial r$ are singular at $r=2m$: in other terms, the singularity $r=2m$ of the standard interval has been "incorporated" in the differentials $du$ and $dv$: this means that, rigorously speaking, eq. (3.1) is *not* a proper extension of the standard form. Finally, at $r=0$, i.e. for $v^2 - u^2 = 1$, the Kruskal-Szekeres solution is singular, but the locus $v^2 - u^2 = 1$ is *space-like*, and therefore cannot represent a *material* structure [4]. Accordingly, eq. (3.1) *does not characterize adequately the concept of mass point*.

**4**. − In 1923 Marcel Brillouin proved in a detailed way a result which is actually rather intuitive, i.e. that Schwarzschild's original form, and the analogous form obtained by substituting (2.3) in (2.1), are **maximally extended** − and that the standard solution is valid **only for** $r>2m$ [5].

The final section of his paper is as follows (I change only the notations): "The conclusion seems to me inescapable: the limit $r=0$ [of the forms obtained with the choices (2.2), (2.3) ] is insurmountable; it embodies the *material* singularity. The distance $d$ from this origin ($r=0$) to a point with co-ordinate radius $r$, calculated along a radius vector ($\theta$ = const.; $\varphi$ = const.) is [for the form corresponding to (2.3) ]

$$d = \sqrt{r(r+2m)} + m\ln\frac{r+m+\sqrt{r(r+2m)}}{m} \quad .$$

The ratio between the circumference $2\pi(r + 2m)$ and the [co-ordinate] radius is everywhere larger than $2\pi$; in particular at the origin ($d=0$; $r=0$) this ratio becomes infinite. […]. It is this singularity [$r=0$] that constitutes what physics calls *the material point* […]". (From the English version quoted in [5]).

We can only add that the curvature invariants of the above singularity at $r=0$ have finite values.

**5**. − There is another and very physical proof that Schwarzschild's original form [3] − or the form corresponding to (2.3) − imply the only reasonable definition of the concept of point mass in general relativity

As is well known, in a second basic work [6] Schwarzschild solved the problem of the Einstein field generated by a sphere of an incompressible and homogeneous fluid. Now, if we calculate, through a suitable limit procedure, the field of a mass point from the field of the fluid, we find again the result (2.1) − (2.2), see [7]. This is clearly a convincing demonstration of the *physical* adequacy of Schwarzschild's *original* concept of material point.

> "Man kann nicht immer zusammen stehn,
> Am wenigsten mit großen Haufen."
> J.W. v. Goethe

APPENDIX

**On space-time singularities**

The space-time singularities can be classified in two different ways: according to mathematical or physical criteria.

The mathematical classifications are scarcely interesting for physical aims, and therefore I shall not discuss them. *Physically*, we have true (physical) and false (nonphysical) singularities. The true singularities are time-like, or light-like, loci in which mass-energy is present.

*Examples*. − The singularity $r=2m$ of the standard form [ $f(r) \equiv r$ in (2.1)] is a nonphysical singularity. The singularity $v^2 - u^2 = 1$ of the Kruskal-Szekeres form [1] is nonphysical because space-like. The singularities $r=0$ of Schwarzschild's original form [3] and of the form investigated by Brillouin [5] are physical singularities: the loci $r=\varepsilon$, with $\varepsilon$ arbitrarily small, are time-like − and *the correspondence with Newton's theory* tells us that there is actually matter at $r=0$.

I emphasize that the Kruskal-Szekeres solution and the solutions of Schwarzschild − Brillouin belong to two *different* pseudo-Riemannian manifolds. −

*A last remark*. The so-called "trapped surfaces" are geometrical loci subordinate to the existence of nonphysical singularities analogous to the surface $r=2m$ of the standard solution. Consequently, theorems on gravitational collapse and cosmological models which utilize in an essential way the notion of trapped surface are wholly void of physical sense.